\documentclass[11pt]{article}

\usepackage[margin=1in]{geometry}
\usepackage{tabularx}
\usepackage{authblk}
\usepackage{needspace}
\usepackage{amsmath, amssymb, amsthm, mathtools}
\usepackage{bbm}
\usepackage{booktabs}
\usepackage{array}
\usepackage{enumitem}
\usepackage{graphicx}
\usepackage{xcolor}
\usepackage{setspace}
\usepackage[numbers]{natbib}
\usepackage[hidelinks]{hyperref}
\usepackage[nameinlink,capitalize,noabbrev]{cleveref}
\usepackage[suppress]{color-edits}

\addauthor{am}{red}

\newtheorem{theorem}{Theorem}[section]
\newtheorem{proposition}[theorem]{Proposition}
\newtheorem{lemma}[theorem]{Lemma}
\newtheorem{corollary}[theorem]{Corollary}

\theoremstyle{definition}

\newtheorem{assumption}[theorem]{Assumption}

\theoremstyle{remark}

\title{%
    \textbf{One-Sided Randomization in Two-Sided Platform Experiments: Which Side Should be Randomized?}
}

\author[1]{Ramesh Johari}
\author[1]{Anushka Murthy}
\affil[1]{Management Science and Engineering, Stanford University}
\date{\today}

\begin{document}

\maketitle

\begin{abstract}
Two-sided marketplace platforms often use randomized experiments
to evaluate new features, raising the question of whether to randomize
customers or listings. Randomizing the smaller (\emph{short}) side
can reduce interference bias due to weaker competition, while randomizing the larger
(\emph{long}) side provides a larger nominal sample, potentially
reducing variance. On the surface, these competing considerations suggest a
bias-variance trade-off in choosing which side to randomize.
We investigate this trade-off in a two-sided marketplace with
first-come, first-served (FCFS) booking and heterogeneous customer
preferences, studying the bias and variance of difference-in-means
estimators of the global average treatment effect.

Our findings suggest that rather than a bias-variance tradeoff, in fact short-side randomization is generally preferable on {\em both} dimensions.  We prove these results in an asymptotic regime where market imbalance increases.  We establish conditions under which (1) short-side randomization eliminates first-order interference bias, while long-side randomization retains it; and (2) short-side randomization also achieves a variance rate of no larger asymptotic order than long-side randomization. The key insight is that although the long side contains more participants, the number of matches is constrained by the population on the short side, limiting the apparent variance advantage of randomizing a larger sample. Our conditions accommodate heterogeneous preferences and treatment effects and are satisfied by familiar logit models of customer choice. Together, these findings support randomizing on the short side, despite the larger sample size available on the long side.
\end{abstract}

\section{Introduction}
\label{sec:intro}

Two-sided marketplace platforms routinely use randomized experiments (A/B tests) to evaluate new interventions and features: e.g., a new design for product listings, or a new search algorithm. The goal of such an experiment is typically to estimate the GATE ({\em global average treatment effect}) with respect to the number of matches. This is the difference in the number of matches between two counterfactual universes: one where the entire market is in the treatment condition ({\em global treatment}), and another where the entire market is in the control condition ({\em global control}). A common procedure to estimate the GATE is to randomize one side of the market (i.e. customers or listings) to treatment or control using simple i.i.d. Bernoulli randomization; collect the match outcomes after the experiment; and form a difference-in-means (DM) estimator for the GATE.

Because these platforms are two-sided, they face a choice: should they randomize on the demand side of the market (``customers''), or on the supply side of the market (``listings'')?  We begin by noting that regardless of which side of the market is chosen, typical approaches to estimation (such as the use of the DM estimator) exhibit a well-known {\em interference bias} \citep{blake2014marketplace,johari2022experimental}, since both treated and control participants on one side interact with the same set of participants on the opposite side. 

\amedit{Two competing considerations inform the choice of which side to randomize; both depend centrally on which side of the market is more constrained.} To fix terminology, we refer to the side of the market with fewer participants as the {\em short} side of the market, and the opposing side as the {\em long} side. One approach to making this choice relies on the intuition that the sample size is larger on the long side, and thus (naively) variance is lower by randomizing on the long side; this is a common view in practice, and leads to long-side randomization.  On the other hand, prior work \citep{johari2022experimental} establishes in a Markovian queueing model of a two-sided marketplace that the bias is lower when randomizing on the short-side. These effects seem in tension with each other, leaving practical guidance unclear.

To study this apparent bias–variance trade-off, we consider a model of randomized experiments in a two-sided marketplace with one-sided customer preferences and first-come, first-served (FCFS) allocation of listings. Customers arrive sequentially and may book one listing from available inventory according to potentially heterogeneous preferences, with each listing accommodating at most one customer. Our model captures marketplace settings in which customers select from available inventory and listings become unavailable once booked, as in many lodging and e-commerce platforms. We allow treatment to affect customer choice and consider two experiment designs: randomizing customers to treatment and control; or randomizing listings to treatment and control. For each design, we study the interference bias and variance of the difference-in-means estimator of the GATE. Our framework accommodates a broader class of customer choice models beyond the logit choice model studied in \cite{johari2022experimental}. 

We focus on large, imbalanced marketplaces, studying sequences of markets in which both sides grow but the ratio of short-side to long-side participants approaches zero. Let $s_n$ denote the fraction of market participants on the short side, and let $n$ denote the total market size. Because the number of matches is bounded by the short-side population, the GATE is always at most of order $s_n$. We therefore investigate whether interference bias remains at first order, $\Theta(s_n)$, or whether the first-order bias vanishes, $o(s_n)$. We also compare the asymptotic variance rates of the two experiment designs, studying conditions under which both achieve variance of order $O(s_n/n)$ despite the long-side having a larger sample.

Our first set of results establishes two sufficient conditions under which short-side randomization eliminates first-order interference bias --- i.e., the bias is $o(s_n)$. The first condition, which we call {\em near saturation}, is that nearly all short-side participants are matched under both global treatment and control, as well as in the experiment. This condition arises naturally when listings are the short side, since in our model, listings do not hold preferences or reject bookings. The second condition applies when customers are the short side and does not require near saturation. Instead, we assume that customers' preferred listings are not overly concentrated on a small subset of available listing inventory. Under this condition, the effect of competition on match outcomes is second-order, yielding a bias of \(O(s_n^2)\) up to an exponentially small remainder.

In contrast, we show that long-side randomization can retain first-order interference bias, $\Theta(s_n)$. We show that this bias arises because treatment can redistribute matches between treated and control participants even when its effect on the total number of matches is small. We then show that the consideration-set logit choice model of \cite{johari2022experimental} satisfies the conditions of our short-side and long-side bias results, thereby extending their conclusions to a broader class of customer choice models.

We next turn to the variance of the two experiment designs. Although randomizing on the long side provides a larger nominal sample size, the number of matches is limited by the short-side population. Consequently, participants on the long side become increasingly unlikely to match as the market becomes more imbalanced, potentially offsetting the variance reduction associated with a larger sample. We formalize this intuition and show that the variance of the difference-in-means estimator is $O(s_n/n)$ under both short-side and long-side randomization.

The remaining question is then whether long-side randomization can still produce a smaller variance of order $o(s_n/n)$. In this direction, we establish sufficient conditions under which long-side randomization has variance $\Theta(s_n/n)$, implying that short-side randomization has no larger-order asymptotic variance. When customers are scarce, we strengthen the bounded-concentration condition used in our bias analysis to hold throughout the booking process, and additionally require that both experiment groups attract nonvanishing demand. When listings are scarce, we instead require that treatment does not change the probability of booking any particular listing by more than a constant multiplicative factor, and that a nonvanishing fraction of listings are booked in expectation. These conditions ensure that there is sufficient uncertainty in how matches are distributed between the treatment and control groups. We again verify that these conditions hold in the consideration-set logit model of \cite{johari2022experimental}.

Our model and results thus demonstrate there is no tradeoff between bias and variance: short-side randomization has a lower-order asymptotic interference bias than long-side randomization, and an asymptotic variance rate of at most the same order.  As noted above, our model is well-matched to many online platform settings with depletable inventory.  Further the relevant conditions we use to establish our main results are relatively mild, and likely to be satisfied in most real-world platforms: they rule out highly concentrated customer preferences, and nearly deterministic allocation of matches between the treatment and control groups.  These conditions accommodate heterogeneous preferences and treatment effects and are satisfied by familiar logit models of customer choice.  For practitioners, our findings suggest that from a statistical standpoint, they should prefer randomizing on the short-side of the market.

\paragraph{Organization of the paper.} In Section \ref{sec:model} we introduce our model of experimentation in a two-sided market with one-sided preferences and first-come, first-served search. In Section \ref{sec:bias} we present our bias results. In Section \ref{sec:fcfs-variance} we present our variance results. The proofs of the bias and variance results appear in Appendices \ref{app:fcfs-bias} and \ref{app:fcfs-variance} respectively.

\subsection{Related Literature}

Our paper contributes to the literature on experiment design in two-sided
marketplaces, where competition for scarce matching opportunities creates
interference between treatment and control participants. Prior work has
studied the resulting estimation bias and developed approaches to mitigating
interference in marketplace experiments
\citep{blake2014marketplace,basse2016randomization,fradkin2019simulation,
holtz2025reducing,liu2021trustworthy}.
Related work studies the estimation of treatment effects in market equilibrium
\citep{wager2021experimenting,munro2025treatment,munro2025causal}.

The papers most closely related to ours study interference bias and variance in conventional one-sided marketplace experiments. In \cite{johari2022experimental}, the authors analyze customer- and listing-randomized experiments in a Markovian marketplace model and shows that randomizing the short side can reduce interference bias. Similarly, \cite{li2022interference} studies the bias and variance of these designs in a stylized matching model and provides guidance on the choice of randomized side. We complement this work by analyzing sequential first-come, first-served matching with general, potentially heterogeneous customer preferences. Rather than imposing a particular parametric choice model or examining mean-field limits, we develop finite-market conditions under which randomizing the short side eliminates first-order interference bias, and conditions under which the variance rate of short-side randomization is no worse than that of long-side randomization. Our results thus address both components of the apparent bias--variance tradeoff in a sequential marketplace with indivisible inventory. Related work on marketplace estimators includes approaches based on shadow prices and optimization \citep{bright2025reducing}, inference under two-sided market randomization \citep{liu2026randomization}, and regression adjustment for double-randomization designs \citep{sudijono2026regression}.

A complementary literature develops alternative experiment designs and estimators to mitigate interference. Two-sided and multiple-randomization designs exploit variation on both sides of a marketplace \citep{johari2022experimental,bajari2023experimental,masoero2026multiple}, and  \cite{chawla2026double} proposes a double-diagonal estimator that uses cross-arm interactions to reduce bias under two-sided randomization. In another direction, \cite{wang2026matching} develops a randomization design for comparing predetermined matching plans using their alternating-path and cycle structure. Other approaches include clustered and network-based designs \citep{ugander2013graph,saveski2017detecting,pougetabadie2019variance,zhao2024clustering}, and marketplace-specific ranking designs \citep{habibi2026ranking}. Other work \citep{shen2026isolation} studies the cost of separating treatment and control into isolated submarkets, which can reduce interference at the expense of thinning matching opportunities. Work on pricing experiments further develops alternative estimators and designs under marketplace congestion \citep{dhaouadi2025price,delarue2025pricing}. In contrast to these approaches, we retain one-sided Bernoulli randomization and the standard difference-in-means estimator, and ask which side of the marketplace should be randomized.

Our analysis also relates to experimentation in dynamic systems, where
treatment assignments can affect future system states and outcomes.
Prior work studies estimation under Markovian models of interference
\citep{farias2022markovian,li2026experimenting,johari2025truncated},
as well as experiment design and inference under temporal interference
using switchback experiments
\citep{bojinov2023switchback,blanchet2026conservative,ni2025decision}.
Related work considers optimal treatment targeting in dynamic systems
\citep{hu2025targeting}. A complementary literature studies the implications of interference
for experiment decision-making. In particular, \cite{johari2024when} examines these impacts on
hypothesis testing and launch decisions in inventory-constrained
marketplaces, while \cite{li2025bandit} studies the selection of
recommendation algorithms when shared learning generates interference.
Our focus, in contrast, is on the bias and variance of the
difference-in-means estimator arising from the choice of which side
of a two-sided marketplace to randomize.

Finally, our work relates to the broader causal-inference literature on interference, including foundational frameworks for treatment effects under interference \citep{hudgens2008toward,aronow2017estimating}, network randomization inference \citep{athey2018exact}, and bipartite exposure mappings \citep{zigler2021bipartite}. Recent work considers robust estimation when interference is only weakly restricted \citep{delarue2026minimax}, and estimation or validation using network structure or temporal outcome information \citep{shirani2025validate,tan2026validating}.  Our work similarly exploits matching structure to obtain an understanding of the bias and variance of one-sided randomization.

\section{Model and Experiment Design}
\label{sec:model}

We consider a two-sided marketplace in which customers arrive sequentially
and choose among available listings. A customer can book at most one listing,
and each listing can accommodate at most one customer. We use this model to
compare two common ways of running an A/B experiment: randomizing customers
or randomizing listings. 

In this section, we introduce the market, the booking process, \amedit{the estimand, the two
experiment designs,} and the estimator. 

\subsection{Market and customer preferences}
\label{subsec:market-preferences}

\paragraph{Customers and listings.}
There are two finite sets of participants: a set of customers
$\mathcal C_n$ with size $N_C$, and a set of listings $\mathcal P_n$ with
size $N_P$. Let $n=N_C+N_P$ denote the total market size. We assume preferences are one-sided, customers hold preferences over listings but each listing would accept any request to book it. Each customer can
complete at most one booking, and each listing has unit capacity.

We study matching over a fixed time period, which we refer to as a
\emph{matching cycle}, following \citet{bright2025reducing}.
Unlike their centralized matching model, customers in our setting
book listings sequentially as they arrive. Bookings are final within
the matching cycle, so booked listings do not become available again,
and customers who leave without booking do not return.

\paragraph{Customer preferences.}
\amedit{For each pair consisting of a customer
$i\in\mathcal C_n$ and a listing $j\in\mathcal P_n$, we assume each pair is given a treatment assignment $w_{ij} \in \{0,1\}$, where $w_{ij}=1$
denotes treatment and $w_{ij}=0$ denotes control. Customer $i$'s utility for booking listing $j$ is then given by
\[
    u_{ij}(w_{ij}).
\]
Note that this utility depends on the pair's treatment assignment.} We normalize the utility of not booking to zero.
Thus, customer $i$ regards listing $j$ as acceptable under treatment $w_{ij}$ if
$u_{ij}(w_{ij})>0$. Among acceptable listings, the customer prefers the listing
with the highest utility. If multiple listings have the same utility, ties are broken
according to a fixed rule.

The intervention may change which listings customers find acceptable,
how customers rank acceptable listings, or both. We allow heterogeneous preferences across customers and listings without imposing a particular choice model. In particular, our framework accommodates the consideration-set logit model of \citet{johari2022experimental}, but does not
require logit choice.

\paragraph{Random market primitives.}
We allow utilities to depend on random customer and listing characteristics.
For each customer $i$, let $\xi_i$ denote a customer-specific random primitive;
for each listing $j$, let $\zeta_j$ denote a listing-specific random primitive.
We assume that the collection
\[
    \bigl(\xi_i:i\in\mathcal C_n\bigr),
    \qquad
    \bigl(\zeta_j:j\in\mathcal P_n\bigr)
\]
consists of mutually independent blocks, and that
\begin{equation}
    u_{ij}(w)=g_{ij,w}(\xi_i,\zeta_j),
    \qquad w\in\{0,1\}.
    \label{eq:fcfs-local-utilities}
\end{equation}

Customers and listings can have heterogeneous characteristics.
A customer's random primitives may affect their preferences for multiple
listings, while a listing's random primitives may affect how multiple
customers value that listing. Thus, preferences may be correlated
across listings for a given customer and across customers for a
given listing. These individual-level random primitives are independent of treatment
assignment.

\subsection{Customer-side sequential booking}
\label{subsec:fcfs-booking}

\paragraph{FCFS booking and available listings.}
Customers arrive in an exogenous order, indexed $1,\ldots,N_C$ for
convenience. When customer $i$
arrives, the customer observes the listings that remain available and
chooses the highest-ranked acceptable listing among them. If none are
acceptable, the customer leaves without booking.

More formally, let $\mathbf w=(w_{ij})$ specify the treatment exposure for
every customer--listing pair, and let $\mathcal A_i(\mathbf w)$ denote the
listings available immediately before customer $i$ arrives. Initially,
$\mathcal A_1(\mathbf w)=\mathcal P_n$. If the available set is nonempty and
\[
    \max_{j\in\mathcal A_i(\mathbf w)}u_{ij}(w_{ij})>0,
\]
customer $i$ books a listing attaining this maximum, using the fixed
tie-breaking rule. That listing is removed from the available set before
the next customer arrives. Otherwise, the available set is unchanged. The
same procedure is applied to every customer in arrival order.

We refer to this allocation rule as \emph{first-come, first-served} (FCFS)
booking. It allows customers to substitute to lower-ranked listings when
their preferred listings have already been booked. The relative ranking of
the remaining listings is unchanged when an unavailable listing is removed.

\paragraph{Match outcomes.}
Let $M_n(\mathbf w)$ denote the matching that results from the FCFS
booking process under treatment exposure vector $\mathbf w$, given
the realized customer and listing primitives. For every participant
$v\in\mathcal C_n\cup\mathcal P_n$, define
\[
    Y_v(\mathbf w)
    =\mathbf 1\{v\text{ is matched in }M_n(\mathbf w)\}.
\]
The total number of matches is
\begin{equation}
    K_n(\mathbf w)
    =\sum_{i\in\mathcal C_n}Y_i(\mathbf w)
    =\sum_{j\in\mathcal P_n}Y_j(\mathbf w).
    \label{eq:fcfs-transaction-count}
\end{equation}
Each booking is counted once on each side of the market. Our outcome of
interest is the number of matches, rather than the identity or value
of the selected customer--listing pairs.

The unit-capacity constraint creates interference between participants.
For example, treating one customer may change the listing they book, which
can change the options available to later customers. Similarly, changing
the treatment status of one listing may alter which customer books it and
the subsequent allocation of other listings.

\subsection{Estimand: Global average treatment effect}
\label{sec:estimand}

\amedit{Our estimand is the difference in expected matches between the global
treatment market, where $\mathbf{w}=\mathbf{1}$, and global control market, where $\mathbf{w}=\mathbf{0}$. We normalize this difference by the total number of
participants and refer to the resulting object as the global average treatment effect (GATE):}
\begin{equation}
    \tau_n
    =\frac{1}{n}\left(
        \mathbb E[K_n(\mathbf{1})]-\mathbb E[K_n(\mathbf{0})]
    \right).
    \label{eq:fcfs-gate}
\end{equation}
Expectations in \eqref{eq:fcfs-gate} are taken over the random market primitives.

\subsection{Experiment designs}
\label{subsec:fcfs-experiments}

\paragraph{Binary treatment.}
For each customer--listing pair $(i,j)$, treatment exposure
$w_{ij}\in\{0,1\}$ determines whether the customer evaluates the listing
under control, with utility $u_{ij}(0)$, or treatment, with utility
$u_{ij}(1)$. We compare two experimental designs that randomize
participants on either the customer side or the listing side.

Let $\pi_C$ and $\pi_P$ denote the treatment assignment probabilities
under customer and listing randomization, respectively. We assume
that for some $\varepsilon>0$,
\begin{equation}
    \varepsilon\leq\pi_d\leq1-\varepsilon,
    \qquad d\in\{C,P\}.
    \label{eq:fcfs-overlap}
\end{equation}
Treatment assignments are independent across participants and
independent of the market primitives.

\paragraph{Customer randomization.}
Under a customer-randomized experiment, each customer
$i\in\mathcal C_n$ receives an assignment
\[
    W_i^C\sim\operatorname{Bernoulli}(\pi_C),
\]
independently across customers. Every interaction involving customer $i$
has treatment exposure given by that customer's assignment:
\begin{equation}
    w_{ij}^{C}=W_i^C,
    \qquad i\in\mathcal C_n,\ j\in\mathcal P_n.
    \label{eq:fcfs-customer-exposure}
\end{equation}
Thus, a treated customer evaluates all listings under treatment, while a
control customer evaluates all listings under control. Treated and control
customers compete for the same pool of available listings.

\paragraph{Listing randomization.}
Under a listing-randomized experiment, each listing
$j\in\mathcal P_n$ receives an assignment
\[
    W_j^P\sim\operatorname{Bernoulli}(\pi_P),
\]
independently across listings. In this case, treatment exposure is determined by the
listing's assignment:
\begin{equation}
    w_{ij}^{P}=W_j^P,
    \qquad i\in\mathcal C_n,\ j\in\mathcal P_n.
    \label{eq:fcfs-listing-exposure}
\end{equation}
An arriving customer may therefore encounter both treated and control
listings and compare them.

Let $M_n^d$ denote the matching obtained under the experiment that
randomizes side $d$, and let $Y_v^d$ be participant $v$'s booking outcome under the matching. The two experimental designs generally produce different matchings,
even though they evaluate the same underlying intervention.
Under customer randomization, capacity constraints create
interference because bookings change the listings available to
subsequent customers. Under listing randomization, interference
also arises because customers choose between treated and control
listings, so treatment can shift bookings between the two groups.

\subsection{Estimator and market imbalance}
\label{subsec:fcfs-estimation}

\paragraph{Difference-in-means estimator.}
For a side-$d$ experiment, where $d\in\{C,P\}$, let
$\mathcal D_d$ denote the set of participants on the randomized side and $N_d$ its size. For
$w\in\{0,1\}$, define
\begin{equation}
\begin{aligned}
    N_{d,w}
    &=\sum_{v\in\mathcal D_d}\mathbf 1\{W_v^d=w\},\\
    K_{d,w}
    &=\sum_{v\in\mathcal D_d}\mathbf 1\{W_v^d=w\}Y_v^d.
\end{aligned}
    \label{eq:fcfs-arm-counts}
\end{equation}
Thus, $N_{d,w}$ is the number of participants assigned to treatment $w$, while
$K_{d,w}$ is the number of those participants who match.
We study the usual difference-in-means estimator, rescaled by the
randomized side's share of the market:
\begin{equation}
    \widehat\Delta_{d,n}
    =\frac{N_d}{n}\left(
        \frac{K_{d,1}}{N_{d,1}}
        -\frac{K_{d,0}}{N_{d,0}}
    \right).
    \label{eq:fcfs-difference-in-means}
\end{equation}
The factor $N_d/n$ puts the estimators from customer and listing
randomization on the same scale as \eqref{eq:fcfs-gate}. We set
$\widehat\Delta_{d,n}=0$ if either experiment group is empty; under our
Bernoulli design this event has probability
$\pi_d^{N_d}+(1-\pi_d)^{N_d}$ and becomes negligible as $N_d$ grows.

Because participants on one side compete for the same constrained set of participants on the other side, the outcome of a treated participant may depend on the assignments of other participants.
Consequently, \eqref{eq:fcfs-difference-in-means} need not be unbiased for
\eqref{eq:fcfs-gate}. We compare the two experiments using the bias and
variance
\begin{equation}
    B_{d,n}=\mathbb E[\widehat\Delta_{d,n}]-\tau_n,
    \qquad
    V_{d,n}=\operatorname{Var}(\widehat\Delta_{d,n}).
    \label{eq:fcfs-bias-variance}
\end{equation}
Throughout the rest of this paper, these expectations and variances include randomness from treatment assignment and random market primitives.

\paragraph{Market imbalance.}
We study how the bias and variance of the two experiment designs change as the market becomes imbalanced in either direction. Let
\begin{equation}
    m_n=\min\{N_C,N_P\},\qquad
    \ell_n=\max\{N_C,N_P\},\qquad
    s_n=\frac{m_n}{n}.
    \label{eq:fcfs-short-side}
\end{equation}
We call the side with $m_n$ participants the \emph{short side}, and the
other side the \emph{long side}. For the rest of this paper, we use $d \in \{0,1,S,L\}$ to denote the global control, global treatment, short-side randomization, and long-side randomization experiment designs respectively. We examine large
markets with $m_n,\ell_n\to\infty$ and
\[
    \frac{m_n}{\ell_n}\longrightarrow0,
    \qquad\text{equivalently,}\qquad s_n\longrightarrow0.
\]
There are two distinct cases: customers may be scarce, with $N_C=m_n$,
or listings may be scarce, with $N_P=m_n$. We also refer to these markets as \emph{demand-constrained} or \emph{suppy-constrained} respectively. Since
each match uses one participant from each side,
$K_n(\mathbf w)\le m_n$, and thus $|\tau_n|\le s_n$.

We study the bias and variance of these estimators along these sequences
of increasingly large and imbalanced markets. We characterize the asymptotic behavior of
$B_{d,n}$ and $V_{d,n}$ under conditions on the market's search
and booking process. With $s_n=m_n/n$ denoting the short side's share of the market,
we study the bias relative to $s_n$ and the variance relative
to $s_n/n$.

\section{Bias}
\label{sec:bias}

In this section, we study how the bias of the difference-in-means estimator
behaves under each experiment design when one side of the market is much smaller than the other.

We establish two different sufficient conditions under which randomizing the short side
eliminates first-order bias. The first sufficient condition is that almost all short-side
participants are matched; this condition holds for the listing side when the market is supply-constrained as they do not hold preferences. The second condition applies when customers are scarce and
they don't concentrate their preferred choices on the same small set of listings. We then give insight into when long-side randomization retains first-order bias.

Throughout the paper, let superscripts $S$ and $L$ on match counts correspond to short-side and
long-side randomization, respectively. As introduced in Section \ref{sec:estimand}, $K_n(\mathbf{0})$ and
$K_n(\mathbf{1})$ denote match counts under global control and global treatment.

\subsection{Bias under near saturation}
\label{subsec:bias-saturation}

We first consider markets in which almost every participant on the short
side matches. Let $K_n^S$ be the number of matches in
the experiment that randomizes the short side, and recall that
$K_n(\mathbf{0})$ and $K_n(\mathbf{1})$ are the match counts under global control
and global treatment. Since at most $m_n$ matches can occur, the
quantities $m_n-K_n^S$, $m_n-K_n(\mathbf{0})$, and $m_n-K_n(\mathbf{1})$ count the
unmatched participants on the short side in the three regimes.

\begin{proposition}[Short-side bias under near saturation]
\label{prop:fcfs-bias-saturation}
Suppose that, along a sequence of markets with $m_n\to\infty$,
the expected fraction of short-side participants who are matched
converges to one under short-side randomization, global treatment,
and global control. That is,
\begin{equation}
    \frac{\mathbb E[K_n^S]}{m_n}\longrightarrow 1,
    \qquad
    \frac{\mathbb E[K_n(\mathbf w)]}{m_n}\longrightarrow 1,
    \quad \mathbf w\in\{\mathbf 0,\mathbf 1\}.
    \label{eq:fcfs-near-saturation}
\end{equation}
Then the difference-in-means estimator under short-side randomization
satisfies
\[
    B_{S,n}=o(s_n).
\]
\end{proposition}

The intuition is straightforward. If nearly all short-side participants
match, their average booking outcomes are close to one in both experiment groups, while total matches are close to $m_n$ under either global
treatment or control market. Consequently, neither the expected value of the difference-in-means estimator nor the global
treatment effect can be large relative to $s_n$. The proof of Proposition~\ref{prop:fcfs-bias-saturation} is in Appendix~\ref{app:fcfs-bias}. The argument behind the result does
not require a particular customer choice model; it uses only the capacity constraints and binary booking outcomes.

\paragraph{Application when listings are scarce.}
Near saturation is natural in the FCFS model when listings are the short
side and customers are abundant. Listings do not have preferences or reject
bookings, so if every customer finds every listing acceptable, all listings
are booked after the first $m_n$ customers arrive. 

More generally, it is enough that each listing attracts sufficiently many
customers who would choose it if all listings were available. For an experiment design
$r\in\{0,1,S\}$, let $J_i^r$ be customer $i$'s preferred acceptable listing
when all listings are available, with $J_i^r=\varnothing$ if no listing is
acceptable. Let $\mathcal E_r$ contain the listing primitives and, in
the experiment market, the listing treatment assignments. Conditional on
$\mathcal E_r$, the choices $J_i^r$ are independent across customers.

\begin{proposition}[A sufficient condition for saturation under FCFS]
\label{prop:fcfs-demand-coverage}
Suppose listings are scarce, so $N_P=m_n$ and $N_C=\ell_n$. If there exists
$c>0$ such that for every listing $j$ and
each experiment design $r\in\{0,1,S\}$,
\begin{equation}
    \sum_{i\in\mathcal C_n}
       \Pr(J_i^r=j\mid\mathcal E_r)
    \ge c\frac{\ell_n}{m_n}
    \qquad\text{almost surely},
    \label{eq:fcfs-demand-coverage}
\end{equation}
then
\[
    \max\left\{\mathbb E[m_n-K_n^S],\,
    \mathbb E[m_n-K_n(\mathbf w)]\right\}
    \le m_n\exp\!\left(-c\frac{\ell_n}{m_n}\right)
    \qquad (\mathbf w\in\{\mathbf 0,\mathbf 1\}).
\]
In particular, $B_{S,n}=o(s_n)$ as $m_n/\ell_n\to0$ and $m_n\to\infty$.
\end{proposition}

Condition~\eqref{eq:fcfs-demand-coverage} ensures that each listing has a
growing expected number of customers who would select it as their first
choice. It is sufficient, but not necessary: a listing may also be booked
by customers who substitute away from their preferred options. The proof
uses the fact that if any customer ranks a listing first, that listing must
eventually be booked under FCFS.

\subsection{Bias under bounded concentration of customer choice}
\label{subsec:bias-dispersion}

We next consider the case where customers are scarce. A customer may remain
unmatched if they retain a nonvanishing outside option even when listings are abundant, so the saturation condition from the previous subsection need
not apply. We instead impose a condition limiting how much competition
customers face for the same listings.

For each customer $i$ and treatment $w\in\{0,1\}$, let
$J_i^{(w)}\in\mathcal P_n\cup\{\varnothing\}$ denote the listing that
customer $i$ would choose under treatment $w$ if all listings were available.
We refer to $J_i^{(w)}$ as the customer's \emph{choice in an isolated market}. Let
$\mathcal F_{P,n}$ be the information generated by the listing-specific
primitives $(\zeta_j)_{j\in\mathcal P_n}$.

\begin{assumption}[Bounded concentration of customer choice]
\label{ass:fcfs-bias-dispersion}
There is a constant $D<\infty$, independent of market size, such that,
almost surely,
\begin{equation}
    \Pr(J_i^{(w)}=j\mid\mathcal F_{P,n})
    \le\frac{D}{\ell_n},
    \qquad
    i\in\mathcal C_n,\quad j\in\mathcal P_n,\quad w\in\{0,1\}.
    \label{eq:fcfs-bias-dispersion}
\end{equation}
\end{assumption}

This assumption rules out listings that attract a disproportionately large
fraction of scarce customers. It allows heterogeneous preferences and does
not require the probabilities to be identical across listings. 

\begin{proposition}[Bias under bounded concentration of customer choices]
\label{prop:fcfs-bias-dispersion}
Suppose customers are scarce and
Assumption~\ref{ass:fcfs-bias-dispersion} holds. Then
\begin{equation}
    |B_{S,n}|
    \le C\frac{m_n^2}{n\ell_n}
    +Cs_n e^{-cm_n}
\end{equation}
for constants $C,c>0$ independent of market size.
Consequently,
\[
    B_{S,n}=o(s_n).
\]
\end{proposition}

To understand the result, first consider what each customer would choose
if no other customer were present. Under
Assumption~\ref{ass:fcfs-bias-dispersion}, the expected number of pairs of
customers that would make the same choice is
$O(m_n^2/\ell_n)$. These repeated choices bound the number of matches
lost because of competition. Since outcomes are normalized by the total
market size $n$, the resulting bias is of order
$m_n^2/(n\ell_n)=O(s_n^2)$. This gives us the first term in the upper bound. The second term arises from
our convention of setting the estimator to zero when one experiment group
is empty; this term is exponentially small in $m_n$. Thus the bias is always $o(s_n)$, and when the long-side population grows polynomially with the short-side
population, the exponentially small term is negligible and the
bias is $O(s_n^2)$.

\subsection{Comparison with long-side randomization}
\label{subsec:bias-long-side}

Unlike short-side randomization, long-side randomization can
retain first-order bias even in highly imbalanced markets. To see why, let $K_n^L$ denote the total number of matches under long-side randomization, and let $K_{n,1}^L$ denote the number of these matches involving treated participants.
The bias satisfies \amcomment{I could have done more here and imposed conditions under which the quantity below converges to a limiting expression that is nonzero. However, none of the other results require the finite market quantities to converge to a deterministic limit; they operate by directly bounding finite market quantities. So I decided to just write out this expression and then for logit choice we establish that the liminf remains positive}

\begin{equation}
\frac{B_{L,n}}{s_n}
=
\frac{1}{\pi(1-\pi)}
\left(
\frac{\mathbb E[K_{n,1}^L]}{m_n}
-\pi\frac{\mathbb E[K_n^L]}{m_n}
\right)
-\frac{\tau_n}{s_n}
+O(\ell_n^{-1/2}).
\label{eq:fcfs-long-side-bias-decomposition}
\end{equation}

The first term captures the extent to which treatment changes the allocation of matches between treated and control participants, while the second captures its effect on the total number of matches. Even when the market is highly imbalanced, these effects need not coincide.

When listings are abundant, making treated listings more attractive can shift bookings away from control listings. When customers are abundant, treated customers may book listings that would otherwise have been available to control customers. In both cases, treatment can change how matches are distributed between experimental groups, even when its effect on the total number of matches under global treatment versus global control is small. This discrepancy can generate first-order interference bias.

We next show that this bias persists in the consideration-set logit model of \citet{johari2022experimental} under natural conditions on the treatment effects.

\subsection{An example with logit choice}
\label{subsec:bias-logit-choice}

We conclude this section by showing how the preceding bias results
apply within a familiar choice model. We consider the consideration-set logit model of \citet{johari2022experimental}, allowing customer preferences, listing attractiveness, and
treatment effects to vary across customer--listing pairs.

Suppose customer $i$ independently considers each available listing
$j$ with probability $p_{ij}^{w_{ij}}$, where $w_{ij}\in\{0,1\}$
denotes the treatment exposure of the pair. Conditional on the set
$\mathcal H_i$ of considered available listings, customer $i$
books listing $j\in\mathcal H_i$ with probability
\begin{equation}
    \frac{v_{ij}^{w_{ij}}}
    {\eta_i N_P+
    \sum_{k\in\mathcal H_i}v_{ik}^{w_{ik}}},
    \label{eq:fcfs-logit-choice}
\end{equation}
where $v_{ij}^w>0$ is the utility weight of listing $j$ for
customer $i$ under exposure $w$, and $\eta_i N_P>0$ is the
outside-option weight. The remaining probability is assigned to
not booking.

We impose the following bounds on the model parameters.
There exist positive constants
$\underline p,\underline v,\overline v,
\underline\eta,\overline\eta$, independent of market size, such
that for every customer $i$, listing $j$, and treatment $w$,
\begin{equation}
    \underline p\le p_{ij}^w\le1,
    \qquad
    \underline v\le v_{ij}^w\le\overline v,
    \qquad
    \underline\eta\le\eta_i\le\overline\eta.
    \label{eq:fcfs-logit-heterogeneous-bounds}
\end{equation}

These conditions are similar to those in \citet{johari2022experimental},
who assume strictly positive consideration probabilities and utility
weights, as well as an outside-option weight that scales linearly with
the number of listings. In their finite-type model, positivity implies
uniform bounds on consideration probabilities and utility weights.
We impose these bounds directly to accommodate more general
heterogeneity across customers and listings as the market grows.

To establish positive first-order bias under long-side randomization,
we additionally consider treatments that increase the expected
attractiveness of every customer--listing pair. In particular,
if we write $c_{ij}^w=p_{ij}^wv_{ij}^w$, suppose there exists
$\gamma>0$ independent of market size such that
\begin{equation}
    c_{ij}^1-c_{ij}^0\ge\gamma
    \qquad\text{for all }i,j.
    \label{eq:fcfs-logit-positive-treatment}
\end{equation}
This condition is similar to the sign-consistency assumption of \citet{johari2024when}; it allows heterogeneous treatment effects across customer--listing pairs but requires them to be uniformly positive.

\begin{proposition}[Bias under logit choice]
\label{prop:fcfs-long-side-logit}
Suppose customer choice follows \eqref{eq:fcfs-logit-choice},
the parameter bounds in
\eqref{eq:fcfs-logit-heterogeneous-bounds} hold,
and the treatment probability is bounded away from zero and one.
Along any sequence with $m_n\to\infty$ and
$m_n/\ell_n\to0$, the following hold.

\begin{enumerate}
    \item When either side of the market is scarce, short-side randomization
    eliminates first-order bias:
    \begin{equation}
        B_{S,n}=o(s_n).
    \end{equation}
    Moreover, when customers are scarce,
    \begin{equation}
        |B_{S,n}|
        =O(s_n^2+s_ne^{-c m_n})
    \end{equation}
    for some constant $c>0$.

    \item If the sign-consistency condition
    \eqref{eq:fcfs-logit-positive-treatment} also holds, then
   when either side is scarce, long-side randomization
    retains positive first-order bias:
    \begin{equation}
        \liminf_{n\to\infty}
        \frac{B_{L,n}}{s_n}>0,
        \qquad
        B_{L,n}=\Theta(s_n).
    \end{equation}
\end{enumerate}
\end{proposition}

The two conclusions follow from different properties of the
choice model. When customers are scarce, the parameter bounds
ensure that customers are not too likely to choose the same listings,
so the quadratic bias result for short-side randomization applies.
When listings are scarce, the bounds ensure that nearly all
listings are eventually booked, eliminating first-order
short-side bias.

The additional sign-consistency condition ensures that
long-side randomization retains first-order bias in both cases.
When customers are scarce, treated listings attract bookings
that would otherwise go to control listings. When listings are
scarce, treated customers are more likely to book the remaining
inventory.

\section{Variance}
\label{sec:fcfs-variance}

Having established the bias results, we now show that in a capacity-constrained marketplace, randomizing on the long side need not reduce variance. Regardless of which side is randomized, the number of matches is limited by the size $m_n$ of the short side, so participants on the longer side are less likely to match. To illustrate this, suppose booking outcomes on the randomized side \(d\) were independent, with each participant booking with probability \(p_d\asymp m_n/N_d\). The variance of the rescaled difference-in-means estimator would then be of order

\[
    \left(\frac{N_d}{n}\right)^2
    \frac{p_d(1-p_d)}{N_d}
    =\frac{N_d}{n^2}p_d(1-p_d).
\]

When the long side is randomized, $N_d=\ell_n$ and
$p_d\asymp m_n/\ell_n\to0$, so the variance is of order
$\Theta(s_n/n)$. When the short side is randomized, $N_d=m_n$,
so the variance is at most $O(s_n/n)$. Thus, despite having a
larger sample, long-side randomization need not yield a better
variance rate. We next establish when this conclusion holds under
FCFS booking, where outcomes are dependent.

\subsection{An upper bound for both designs}

Our first result holds under the general FCFS model introduced in Section~\ref{sec:model}, without imposing any additional assumptions on customer preferences.

\begin{proposition}[Variance upper bound]
\label{prop:fcfs-variance-upper}
Under the assumptions of Section~\ref{sec:model}, for either
$d\in\{C,P\}$,
\begin{equation}
    \operatorname{Var}(\widehat\Delta_{d,n})
    =O\left(\frac{m_n}{n^2}\right)
    =O\left(\frac{s_n}{n}\right).
    \label{eq:fcfs-variance-upper}
\end{equation}
\end{proposition}

To understand the result, consider changing the preferences of a single customer or the characteristics of a single listing, while holding everything else fixed. Although this may change which participants are matched to one another, it can change whether at most two customers and two listings are matched. Moreover, if the affected participant remains unmatched both before and after the change, no booking outcomes change. The proof uses these properties, together with the fact that there are at most $m_n$ matches, to establish the variance bound. It also accounts for the random treatment group sizes in the difference-in-means estimator.

\subsection{A lower bound for long-side randomization}

The upper bound alone does not rule out a variance advantage for
long-side randomization. We next give sufficient conditions under which
its variance is at least of order $s_n/n$. The conditions depend on which
side is scarce.

\paragraph{Customers are scarce.}
Suppose $N_C=m_n$ and $N_P=\ell_n$, so the long-side experiment
randomizes listings. Let $\mathcal E_n$ contain the listing primitives and
listing treatment assignments, but not the random customer primitives.
For an available set $A\subseteq\mathcal P_n$, let $J_i(A)$ be the
listing that customer $i$ would choose from $A$, with
$J_i(A)=\varnothing$ if the customer would not book.

\begin{assumption}[Bounded choice probabilities and sufficient demand for both experiment groups]
\label{ass:fcfs-variance-customers-scarce}
There are constants $D<\infty$, $\alpha>0$, and $\eta>0$ such that:
\begin{enumerate}
    \item For every customer $i$, every available set $A$ with
    $|\mathcal P_n\setminus A|\le m_n$, and every $j\in A$,
    \begin{equation}
        \Pr(J_i(A)=j\mid\mathcal E_n)
        \le\frac{D}{\ell_n}
        \quad\text{almost surely}.
        \label{eq:fcfs-variance-uniform-dispersion}
    \end{equation}
    \item With probability tending to one, at least $\alpha m_n$
    customers satisfy both
    \begin{equation}
    \begin{aligned}
        \Pr\bigl(J_i(\mathcal P_n)\in\mathcal P_{n,1}
             \mid\mathcal E_n\bigr)&\ge\eta,\\
        \Pr\bigl(J_i(\mathcal P_n)\in\mathcal P_{n,0}
             \mid\mathcal E_n\bigr)&\ge\eta,
    \end{aligned}
    \label{eq:fcfs-variance-choice-overlap}
    \end{equation}
    where $\mathcal P_{n,w}$ is the set of listings assigned to treatment $w$.
\end{enumerate}
\end{assumption}

The first condition requires that when customers are scarce, they
are unlikely to choose the same listings, even after earlier bookings
have removed some listings. The second condition ensures that both experiment groups attract
sufficient potential demand: a nonvanishing fraction of customers have
a meaningful probability of choosing a listing from either group.
Together, these conditions ensure that competition does not substantially
reduce the variation arising from customers' choices between the two
experiment groups. 

\paragraph{Listings are scarce.}
Now suppose $N_P=m_n$ and $N_C=\ell_n$, so the long-side experiment
randomizes customers. Conditional on the listing primitives, write
\[
    Q_i^w(j\mid A)
    :=\Pr\bigl(\text{customer }i\text{ books }j
         \mid A\text{ available},\ W_i^C=w,
         \text{ listing primitives}\bigr)
\]
for $j\in A$, and define $Q_i^w(\varnothing\mid A)$ analogously.
These probabilities may depend on the customer, the particular listing,
and the full available set.

\begin{assumption}[Treatment has bounded impact on booking]
\label{ass:fcfs-variance-listings-scarce}
There are constants $1\le M<\infty$ and $\kappa>0$ independent of
market size such that, conditional on the listing primitives, for every
customer $i$, every reachable available set $A$, and every $j\in A$,
\begin{equation}
    M^{-1}Q_i^0(j\mid A)
    \le Q_i^1(j\mid A)
    \le M Q_i^0(j\mid A)
    \quad\text{almost surely},
    \label{eq:fcfs-variance-booking-overlap}
\end{equation}
and the expected number of matches in the customer-randomized
experiment satisfies
\begin{equation}
    \mathbb E[K_n^L]\ge\kappa m_n.
    \label{eq:fcfs-variance-positive-utilization}
\end{equation}
\end{assumption}

The first condition requires a booking of a particular listing to
remain possible under both customer treatment assignments with comparable
probabilities. The condition allows customers and listings to differ in their
characteristics and preferences, as well as in how treatment affects
their choices. The second condition requires that the expected number
of matched listings is at least a constant fraction of all listings.

Under these conditions, we can show that randomizing on the long side retains substantial variance. The intuition is that even after observing which customers successfully
book, there is uncertainty about their treatment assignments which generates meaningful variance in the estimator.

\begin{theorem}[Variance under long-side randomization]
\label{thm:fcfs-variance-long-lower}
Suppose $m_n,\ell_n\to\infty$ and $m_n/\ell_n\to0$. If customers
are scarce, impose Assumption~\ref{ass:fcfs-variance-customers-scarce}.
If listings are scarce, impose
Assumption~\ref{ass:fcfs-variance-listings-scarce}.
Then
\begin{equation}
    \operatorname{Var}(\widehat\Delta_{L,n})
    =\Theta\left(\frac{m_n}{n^2}\right)
    =\Theta\left(\frac{s_n}{n}\right).
    \label{eq:fcfs-variance-long-lower}
\end{equation}
Consequently,
\[
    \operatorname{Var}(\widehat\Delta_{S,n})
    =O\bigl(\operatorname{Var}(\widehat\Delta_{L,n})\bigr).
\]
\end{theorem}

The intuition behind these results is that although the long side
contains more participants, the number of matches is limited by
the size of the short side. Consequently, the larger sample does not
necessarily translate into a lower variance rate. The key is to show
that there remains sufficient variation in how matches are
distributed between treatment and control. When customers are scarce,
customers are not overly likely to choose the same listings, so competition
does not substantially reduce this variation. When listings are
scarce, customers have comparable probabilities of booking a given
listing under treatment and control, so observing which customers
successfully book does not fully reveal their treatment assignments.

\subsection{Variance comparison under logit choice}

We now apply our variance results to the consideration-set logit
model introduced in Section~\ref{subsec:bias-logit-choice}.
The same logit parameter bounds in \eqref{eq:fcfs-logit-heterogeneous-bounds} that guarantee vanishing first-order
bias under short-side randomization also ensure that long-side randomization cannot achieve a smaller asymptotic variance.

\begin{corollary}[Variance comparison under logit choice]
\label{cor:fcfs-variance-logit}
Suppose customer choice follows \eqref{eq:fcfs-logit-choice},
the parameter bounds in
\eqref{eq:fcfs-logit-heterogeneous-bounds} hold,
and the treatment probabilities are bounded away from zero
and one. Then, when either customers or listings are scarce,
\[
    \operatorname{Var}(\widehat\Delta_{S,n})
    =O\left(\frac{s_n}{n}\right),
    \qquad
    \operatorname{Var}(\widehat\Delta_{L,n})
    =\Theta\left(\frac{s_n}{n}\right).
\]
\end{corollary}

When customers are scarce, the parameter bounds ensure that
customers are not overly likely to compete for the same listings, while
both treated and control listings retain a positive probability
of being chosen. When listings are scarce, the bounds ensure
that treatment and control customers have comparable probabilities
of booking any available listing. Consequently, the larger
nominal sample under long-side randomization does not yield a smaller asymptotic variance rate.

Together with Proposition~\ref{prop:fcfs-long-side-logit},
this corollary shows that short-side randomization eliminates
first-order bias without having a worse variance rate.

\section{Conclusion}
\label{sec:conclusion}

Our results suggest that platforms should generally favor short-side randomization when designing experiments in two-sided marketplaces. Under relatively mild conditions satisfied by familiar logit models of customer choice, short-side randomization reduces interference bias while achieving an asymptotic variance rate no worse than long-side randomization. This challenges the conventional intuition that randomizing on the larger side necessarily yields lower variance: although the long side provides a larger nominal sample, the number of matches is constrained by the short-side population, limiting the benefit of randomizing more participants. An important direction for future work is to develop practical methods for selecting the randomized side based on observable market characteristics.

\section{Acknowledgments}

This work was supported in part by the National Science Foundation, the Lillie Fund, Stanford Data Science, and Stanford HAI. We are grateful to Guido Imbens and Stefan Wager for helpful conversations and feedback on this work.

\bibliographystyle{apalike}
\bibliography{references}

\clearpage
\appendix

\section{Proofs for Section~\ref{sec:bias}}
\label{app:fcfs-bias}

This appendix proves the bias results in Section~\ref{sec:bias}. The main
ingredient is a simple property of sequential booking: a listing that is
someone's first acceptable choice when all listings are available must be
booked in the full market. This observation allows us to bound the effect
of competition without restricting how customers rank their remaining
options.

\subsection{First choices and match count losses}
\label{app:fcfs-bias-first-choices}

Fix a realization of the market and a vector of treatments $\mathbf w$.
For each customer $i$, let
$J_i\in\mathcal P_n\cup\{\varnothing\}$ be the listing customer $i$
would choose if all listings were available, with $J_i=\varnothing$ if
none is acceptable. Write $A_i=\mathbf 1\{J_i\ne\varnothing\}$ and let
$Y_i$ denote the customer's actual booking indicator under FCFS.

\begin{lemma}[Bounding the number of lost matches]
\label{lem:fcfs-first-choice-coverage}
Let $U=\{J_i:J_i\ne\varnothing\}$ be the set of distinct nonempty
first-choice listings, and let $K$ be the number of matches. Then
$U$ is contained in the set of booked listings and
\begin{equation}
\begin{aligned}
    0
    &\le \sum_{i\in\mathcal C_n}(A_i-Y_i)
      =\sum_{i\in\mathcal C_n}A_i-K \\
    &\le \sum_{i\in\mathcal C_n}A_i-|U| \\
    &\le \sum_{\substack{i,k\in\mathcal C_n\\i<k}}
        \mathbf 1\{J_i=J_k\ne\varnothing\}.
\end{aligned}
    \label{eq:fcfs-first-choice-loss}
\end{equation}
\end{lemma}

\begin{proof}
If customer $i$ has no acceptable listing when all listings are
available, the customer cannot match in the full market. Therefore
$Y_i\le A_i$ for every $i$, proving the first inequality.

Consider any listing $j\in U$. If $j$ has already been booked when a
customer with first choice $j$ arrives, then $j$ is matched in the final
allocation. Otherwise that customer finds $j$ available and selects it,
because $j$ is their highest-ranked acceptable listing. In either case,
$j$ is eventually booked. Hence $K\ge |U|$.

For each listing $j$, let
$d_j=\sum_i\mathbf 1\{J_i=j\}$. We have
\[
    \sum_i A_i-|U|=\sum_{j\in\mathcal P_n} d_j-\sum_{j \in \mathcal P_n}\mathbf{1}_{d_j \geq 1}=\sum_{j\in\mathcal P_n}(d_j-1)_+.
\]
For every nonnegative integer $d$, $(d-1)_+\le\binom d2$. Thus,
\[
\sum_{j\in\mathcal P_n}(d_j-1)_+ \leq \sum_{j\in\mathcal P_n}\binom{d_j}{2}.
\]
The sum over listings gives the final inequality in
\eqref{eq:fcfs-first-choice-loss}, since
$\sum_j\binom{d_j}{2}$ counts pairs of customers with the same nonempty
first choice.
\end{proof}

The lemma above bounds the number of matches lost relative to a market where customers could choose whatever they want regardless of capacity constraints.

\subsection{Near saturation}
\label{app:fcfs-bias-saturation}

\begin{proof}[Proof of Proposition~\ref{prop:fcfs-bias-saturation}]
Write $U^S=m_n-K_n^S$ and $U^{(w)}=m_n-K_n(\mathbf w)$,
$w\in\{0,1\}$, for the numbers of unmatched short-side participants
under short-side randomization and the two global regimes. In the experiment, let
$U_{S,w}=N_{S,w}-K_{S,w}$ be the number unmatched short-side participants in treatment $w$. On the event
that both experiment groups are nonempty,
\begin{equation}
\begin{aligned}
    \widehat\Delta_{S,n}
    &=\frac{m_n}{n}\left(
        \frac{K_{S,1}}{N_{S,1}}-
        \frac{K_{S,0}}{N_{S,0}}\right)\\
    &=\frac{m_n}{n}\left(
        \frac{U_{S,0}}{N_{S,0}}-
        \frac{U_{S,1}}{N_{S,1}}\right).
\end{aligned}
    \label{eq:fcfs-saturation-dm}
\end{equation}

Let $\varepsilon>0$ be the lower bound on treatment and control
probabilities from Section~\ref{subsec:fcfs-experiments}, and define
\[
    \mathcal B_n=
    \left\{N_{S,1}\ge\frac{\varepsilon m_n}{2},\quad
           N_{S,0}\ge\frac{\varepsilon m_n}{2}\right\}.
\]
Bernoulli concentration gives
$\Pr(\mathcal B_n^c)\le2\exp(-\varepsilon m_n/8)$.
On $\mathcal B_n$, equation~\eqref{eq:fcfs-saturation-dm} implies
\[
    |\widehat\Delta_{S,n}|
    \le\frac{2}{\varepsilon n}(U_{S,0}+U_{S,1})
    =\frac{2U^S}{\varepsilon n}.
\]
On $\mathcal B_n^c$, we use 
\[
    |\widehat\Delta_{S,n}|
    \le\frac{m_n}{n}=s_n.
\]
Consequently,
\begin{equation}
    |\mathbb E[\widehat\Delta_{S,n}]|
    \le\frac{2\mathbb E[U^S]}{\varepsilon n}
       +2s_n\exp(-\varepsilon m_n/8).
    \label{eq:fcfs-saturation-dm-bound}
\end{equation}

Under global treatment and control,
\[
    \tau_n=\frac{\mathbb E[U^{(0)}]-\mathbb E[U^{(1)}]}{n},
\]
so
\[
    |\tau_n|
    \le\frac{\mathbb E[U^{(0)}]+\mathbb E[U^{(1)}]}{n}.
\]
By the near-saturation condition \ref{eq:fcfs-near-saturation}, the expected number of unmatched
short-side participants is $o(m_n)$ in the experiment and under both
global treatment and control. Together with the preceding bounds, this
implies that both $\mathbb E[\widehat\Delta_{S,n}]$ and $\tau_n$
are $o(s_n)$. Therefore, $B_{S,n}=o(s_n)$.
\end{proof}

\begin{proof}[Proof of Proposition~\ref{prop:fcfs-demand-coverage}]
Fix a regime $r\in\{0,1,S\}$ and condition on $\mathcal E_r$. Let $K$ denote its
total match count, namely $K_n^S$ if $r=S$, $K_n(\mathbf 0)$ if
$r=0$, and $K_n(\mathbf 1)$ if $r=1$. Under the independent
customer primitives in Section~\ref{subsec:market-preferences}, the
choices $J_i^r$ in the isolated market are conditionally independent. By
Lemma~\ref{lem:fcfs-first-choice-coverage}, if listing $j$ is unmatched,
then no customer has $J_i^r=j$. Thus
\begin{equation}
\begin{aligned}
    \Pr(Y_j^r=0\mid\mathcal E_r)
    &\le \prod_{i\in\mathcal C_n}
           \left(1-\Pr(J_i^r=j\mid\mathcal E_r)\right)\\
    &\le \exp\!\left(-\sum_{i\in\mathcal C_n}
           \Pr(J_i^r=j\mid\mathcal E_r)\right)\\
    &\le \exp\!\left(-c\frac{\ell_n}{m_n}\right).
\end{aligned}
    \label{eq:fcfs-listing-unmatched-probability}
\end{equation}
Since $m_n-K$ is the number of unmatched listings, we can write
\[
    m_n-K
    =\sum_{j\in\mathcal P_n}\mathbf 1\{Y_j^r=0\}.
\]
Taking expectations and applying the law of iterated expectations,
\begin{align*}
    \mathbb E[m_n-K]
    &=\sum_{j\in\mathcal P_n}\Pr(Y_j^r=0)\\
    &=\sum_{j\in\mathcal P_n}
      \mathbb E\!\left[\Pr(Y_j^r=0\mid\mathcal E_r)\right]\\
    &\le m_n\exp\!\left(-c\frac{\ell_n}{m_n}\right),
\end{align*}
where the inequality follows from
\eqref{eq:fcfs-listing-unmatched-probability}.

The right side is $o(m_n)$ when $m_n/\ell_n\to0$, uniformly over the
three regimes. Applying Proposition~\ref{prop:fcfs-bias-saturation} completes the
proof.
\end{proof}

\subsection{Bounded concentration of customer choices}
\label{app:fcfs-bias-dispersion}

\begin{proof}[Proof of Proposition \ref{prop:fcfs-bias-dispersion}]
Throughout this proof customers are the short side, so
$N_C=m_n$ and $N_P=\ell_n$. For $w\in\{0,1\}$, define
$A_i^{(w)}=\mathbf 1\{J_i^{(w)}\ne\varnothing\}$. Let
\begin{equation}
    \widetilde\tau_n
    =\frac1n\sum_{i\in\mathcal C_n}
        \left(\mathbb E[A_i^{(1)}]-\mathbb E[A_i^{(0)}]\right)
    \label{eq:fcfs-isolated-target}
\end{equation}
be the global treatment effect that would arise if every customer could
make their choice in isolation without competing with other customers. Note that in the absence of competition, the resulting difference-in-means estimator would be unbiased.

\paragraph{Step 1: Bounding the effect of competition on bookings.}
Conditional on $\mathcal F_{P,n}$, the random variables
$J_i^{(w)}$ are independent across customers. For distinct $i,k$,
Assumption~\ref{ass:fcfs-bias-dispersion} gives
\begin{equation}
\begin{aligned}
    \Pr(J_i^{(w)}=J_k^{(w)}\ne\varnothing
        \mid\mathcal F_{P,n})
    &=\sum_{j\in\mathcal P_n}
       \Pr(J_i^{(w)}=j\mid\mathcal F_{P,n})
       \Pr(J_k^{(w)}=j\mid\mathcal F_{P,n})\\
    &\le\frac{D}{\ell_n}.
\end{aligned}
    \label{eq:fcfs-pair-collision}
\end{equation}
Lemma~\ref{lem:fcfs-first-choice-coverage} therefore implies
\[
    0\le\sum_i\mathbb E[A_i^{(w)}]-\mathbb E[K_n(\mathbf w)]
    \le\frac{D}{\ell_n}\binom{m_n}{2},
    \qquad w\in\{0,1\}.
\]
Taking the difference between the true GATE and the isolated-market GATE and applying the triangle inequality yields
\begin{equation}
    |\tau_n-\widetilde\tau_n|
    \le\frac{D\,m_n(m_n-1)}{n\ell_n}.
    \label{eq:fcfs-global-isolated-error}
\end{equation}

\paragraph{Step 2: Bounding the effect of competition on the estimator.}
For customer randomization, let $A_i^S=A_i^{(W_i^C)}$ and define
\begin{equation}
    \widetilde\Delta_{S,n}
    =\frac{m_n}{n}\left(
       \frac{\sum_{i:W_i^C=1}A_i^{(1)}}{N_{S,1}}
       -\frac{\sum_{i:W_i^C=0}A_i^{(0)}}{N_{S,0}}
    \right),
    \label{eq:fcfs-isolated-dm}
\end{equation}
with the same convention that if an experiment group is empty then the corresponding term is zero.

Let $\mathcal B_{i-1}$ be the set of listings booked before customer $i$
arrives in the experiment. If $A_i^S=1$ but customer $i$ fails to book,
then their first-choice listing $J_i^{(W_i^C)}$ in the isolated market must belong to
$\mathcal B_{i-1}$; otherwise it would still be available and acceptable.
Conditional on $\mathcal F_{P,n}$ and the full customer assignment vector,
$\mathcal B_{i-1}$ depends only on earlier customers' random inputs and is
independent of $J_i^{(W_i^C)}$. Since
$|\mathcal B_{i-1}|\le i-1$, Assumption \ref{ass:fcfs-bias-dispersion} implies
\begin{equation}
\begin{aligned}
    \mathbb E[A_i^S-Y_i^S\mid
          \mathcal F_{P,n},\mathbf W^C]
    &\le\Pr(J_i^{(W_i^C)}\in\mathcal B_{i-1}
          \mid\mathcal F_{P,n},\mathbf W^C)\\
    &\le\frac{D(i-1)}{\ell_n}.
\end{aligned}
    \label{eq:fcfs-individual-booking-loss}
\end{equation}

When both experiment groups are nonempty, the difference between
$\widetilde\Delta_{S,n}$ and $\widehat\Delta_{S,n}$ is bounded by
the average number of bookings lost to competition in each group. Applying
\eqref{eq:fcfs-individual-booking-loss} for any realized treatment assignment
vector with nonempty groups implies
\begin{equation}
\begin{aligned}
&\mathbb E\bigl[|\widehat\Delta_{S,n}-\widetilde\Delta_{S,n}|
          \mid\mathcal F_{P,n},\mathbf W^C\bigr]\\
&\qquad\le\frac{m_n}{n}
  \left(\frac{1}{N_{S,1}}\sum_{i:W_i^C=1}\frac{D(i-1)}{\ell_n}
       +\frac{1}{N_{S,0}}\sum_{i:W_i^C=0}\frac{D(i-1)}{\ell_n}\right)\\
&\qquad\le\frac{2D\,m_n(m_n-1)}{n\ell_n}.
\end{aligned}
    \label{eq:fcfs-experiment-isolated-error}
\end{equation}
Both estimators are zero when an experiment group is empty, so the same upper bound
holds after averaging over all assignments.

\paragraph{Step 3: Accounting for random experiment group sizes.}
Let
$q_{m_n}=\pi_C^{m_n}+(1-\pi_C)^{m_n}$ be the probability of an empty
experiment group. Conditional on any treatment count
$N_{S,1}=k\in\{1,\ldots,m_n-1\}$, the treated participants form a
uniformly chosen subset of size $k$, independently of the potential
isolated-market outcomes. Consequently, the conditional expectation of
\eqref{eq:fcfs-isolated-dm} is $\widetilde\tau_n$, and therefore
\begin{equation}
    \mathbb E[\widetilde\Delta_{S,n}]
       =(1-q_{m_n})\widetilde\tau_n.
    \label{eq:fcfs-isolated-dm-expectation}
\end{equation}
Since $|\widetilde\tau_n|\le m_n/n=s_n$,
\[
    |\mathbb E[\widetilde\Delta_{S,n}]-\widetilde\tau_n|
    \le s_nq_{m_n}.
\]
Combining this inequality with
\eqref{eq:fcfs-global-isolated-error} and
\eqref{eq:fcfs-experiment-isolated-error} proves
\[
    |B_{S,n}|
    \le\frac{3D\,m_n(m_n-1)}{n\ell_n}+s_nq_{m_n},
\]
as claimed.

Finally,
$m_n^2/(n\ell_n)=s_n(m_n/\ell_n)=O(s_n^2)$ as
$m_n/\ell_n\to0$, and $q_{m_n}$ decreases exponentially when
$m_n\to\infty$. The asserted asymptotic conclusions follow.
\end{proof}

\subsection{Application to logit choice model}

\begin{proof}[Proof of Proposition~\ref{prop:fcfs-long-side-logit}]
Write $m=m_n$, $\ell=\ell_n$, and $s=m/n$, and denote the GATE
by $G_n$. The parameter bounds imply, for every $i$, available
set $A$, exposure vector, and $j\in A$,
\begin{equation}
 \frac{a_-}{N_P}\le\Pr(J_i(A)=j)\le\frac{a_+}{N_P},
 \quad
 a_-:=\frac{\underline p\,\underline v}
            {\overline\eta+\overline v},
 \quad
 a_+:=\frac{\overline v}{\underline\eta}.
 \label{eq:fcfs-logit-bias-choice-bounds}
\end{equation}
When customers are scarce, $N_P=\ell$, so the upper bound
verifies the condition of the earlier quadratic short-side bias
result, which gives
$|B_{S,n}|=O(s^2+se^{-cm})$ for some $c>0$.
When listings are scarce, $N_P=m$, and the lower bound shows
that an available listing survives each customer arrival with
probability at most $1-a_-/m$. Hence every listing remains
unmatched after $\ell$ arrivals with probability at most
$e^{-a_-\ell/m}$, uniformly over treatment assignments.
The earlier saturation argument gives
$|B_{S,n}|\le3se^{-a_-\ell/m}=o(s)$.

Now also impose \eqref{eq:fcfs-logit-positive-treatment}.
For long-side randomization, define
\[
 H_{L,n}:=\frac{K_{n,1}^L}{\pi}
          -\frac{K_{n,0}^L}{1-\pi}.
\]
By Lemma~\ref{lem:fcfs-variance-random-denominators},
\begin{equation}
 \mathbb E|n\widehat\Delta_{L,n}-H_{L,n}|
 =O(m/\sqrt{\ell})=o(m).
 \label{eq:fcfs-logit-bias-denominator-transfer}
\end{equation}

\paragraph{Customers scarce.}
Define
$\bar c_{iw}:=\ell^{-1}\sum_{j=1}^{\ell}c_{ij}^w$
and $D_{i,\pi}:=\pi\bar c_{i1}+(1-\pi)\bar c_{i0}$.
For customer $i$ choosing from the full listing set, let
$T_{iw}$ be the total considered utility weight in listing
treatment $w$. Independence of consideration and listing
assignments across listings gives
$\mathbb E[T_{i1}]=\pi\ell\bar c_{i1}$,
$\mathbb E[T_{i0}]=(1-\pi)\ell\bar c_{i0}$, and
$\operatorname{Var}(T_{iw})=O(\ell)$.
Since the logit denominator is at least
$\underline\eta\ell$, the probabilities of choosing from
the two experiment groups are, respectively,
\[
 \frac{\pi\bar c_{i1}}{\eta_i+D_{i,\pi}}
 +O(\ell^{-1/2}),
 \qquad
 \frac{(1-\pi)\bar c_{i0}}{\eta_i+D_{i,\pi}}
 +O(\ell^{-1/2}),
\]
uniformly over $i$. By
\eqref{eq:fcfs-logit-bias-choice-bounds}, competition
changes expected booking counts by $O(m^2/\ell)=o(m)$.
The same reasoning applies to fully treated and controlled
markets, giving
\begin{align*}
 \frac{\mathbb E[H_{L,n}]}m
 &=\frac1m\sum_{i=1}^m
    \frac{\bar c_{i1}-\bar c_{i0}}
         {\eta_i+D_{i,\pi}}+o(1),\\
 \frac{G_n}s
 &=\frac1m\sum_{i=1}^m
    \frac{\eta_i(\bar c_{i1}-\bar c_{i0})}
         {(\eta_i+\bar c_{i1})(\eta_i+\bar c_{i0})}
    +o(1).
\end{align*}
Using \eqref{eq:fcfs-logit-bias-denominator-transfer},
\begin{equation}
 \frac{B_{L,n}}s
 =\frac1m\sum_{i=1}^m(\bar c_{i1}-\bar c_{i0})
  \left[
    \frac1{\eta_i+D_{i,\pi}}
    -\frac{\eta_i}
      {(\eta_i+\bar c_{i1})(\eta_i+\bar c_{i0})}
  \right]+o(1).
 \label{eq:fcfs-logit-bias-hetero-expansion}
\end{equation}
The first factor is at least $\gamma$. Combining the two
fractions in brackets gives the strictly positive numerator
$\eta_i[(1-\pi)\bar c_{i1}+\pi\bar c_{i0}]
+\bar c_{i1}\bar c_{i0}$.
The uniform parameter bounds therefore imply
$\liminf B_{L,n}/s>0$.

\paragraph{Listings scarce.}
For an available set $A$ of size $k$, define
\[
 S_i^w(A):=\sum_{j\in A}C_{ij}(w)v_{ij}^w,
 \qquad
 \mu_i^w(A):=\mathbb E[S_i^w(A)]
 =\sum_{j\in A}c_{ij}^w.
\]
Write $r_i^w(A)$ for the probability that customer $i$
books from $A$ under exposure $w$. With
$f_i(x):=x/(\eta_i m+x)$, we have
$r_i^w(A)=\mathbb E[f_i(S_i^w(A))]$.
Since $\operatorname{Var}(S_i^w(A))\le k\overline v^2$
and $|f_i''|\le 2/(\underline\eta^2m^2)$,
Taylor's theorem gives
\[
 |r_i^w(A)-f_i(\mu_i^w(A))|
 \le \frac{k\overline v^2}{\underline\eta^2m^2}.
\]
The treatment-effect condition yields
$\mu_i^1(A)-\mu_i^0(A)\ge\gamma k$, and hence
\[
 r_i^1(A)-r_i^0(A)
 \ge\frac{\underline\eta\gamma}
          {(\overline\eta+\overline v)^2}\frac{k}{m}
      -\frac{k\overline v^2}{\underline\eta^2m^2}
 \ge b\frac{k}{m}
\]
for some $b>0$ and all sufficiently large $m$.
Also, $r_i^w(A)\le a_+k/m$.
If $A_i$ is the available set when customer $i$ arrives,
independence of the customer's treatment assignment from $A_i$
gives
\[
 \mathbb E[H_{L,n}]
 =\sum_{i=1}^{\ell}
   \mathbb E[r_i^1(A_i)-r_i^0(A_i)]
 \ge b\sum_{i=1}^{\ell}
   \mathbb E\!\left[\frac{|A_i|}{m}\right]
 \ge \frac{b}{a_+}\mathbb E[K_n^L].
\]
The saturation bound implies
$\mathbb E[K_n^L]=m-o(m)$ and $G_n=o(s)$.
By \eqref{eq:fcfs-logit-bias-denominator-transfer},
$\liminf B_{L,n}/s\ge b/a_+>0$.
Finally, $|H_{L,n}|\le m/\min\{\pi,1-\pi\}$,
so the same denominator bound and $|G_n|\le s$
give $B_{L,n}=O(s)$ in both regimes.
\end{proof}

\section{Proofs for the Variance Results}
\label{app:fcfs-variance}

For $d\in\{C,P\}$, let $\mathcal D_d$ denote the randomized side,
let $N_d=|\mathcal D_d|$, and let $\pi_d$ be its treatment probability.
Recall the experiment group match counts $K_{d,w}$ from
\eqref{eq:fcfs-arm-counts}. Write
\begin{equation}
    H_{d,n}
    :=n\widehat\Delta^{\mathrm{HT}}_{d,n}
    =\frac{K_{d,1}}{\pi_d}
      -\frac{K_{d,0}}{1-\pi_d}
    \label{eq:fcfs-variance-HT}
\end{equation}
for the scaled Horvitz--Thompson (HT) estimator. We first establish two finite-market
lemmas that allow us to apply variance bounds for $H_{d,n}$ to the difference-in-means estimator.

\begin{lemma}[Bounded sensitivity of match statuses]
\label{lem:fcfs-variance-stability}
Consider two FCFS markets with the same arrival order and all
primitives identical except the primitives of one customer or one listing.
The final matched/unmatched indicators differ for at most two customers
and at most two listings. If the changed participant is unmatched in
both markets, the booking choices of every participant are identical.
\end{lemma}

\begin{proof}
We compare markets through a common market in which the changed
participant is removed. All preferences of participants other than the
changed participant over their remaining options are the same in
these comparisons, as required by condition
\eqref{eq:fcfs-local-utilities}.

First suppose that listing $j$ is changed. Compare either version of
the original market with a market in which $j$ is unavailable from
the outset. At every arrival, the available set in the market with $j$
is either the same as, or consists of one additional listing relative
to, the available set without $j$. To see this, initially the
difference is $\{j\}$. If both customers choose the same listing in
the two markets, the difference is unchanged. If their choices
differ, the customer in the market with $j$ must choose its one extra
listing since the rankings over all listings common to both markets agree. The customer in
the market without $j$ either chooses a listing common to both markets (which then
becomes the new extra listing in the larger market) or chooses no
listing (in which case the two available sets become equal and stay
equal). Hence a customer can book in the market with $j $ but not in
the market without $j$ at most once; the reverse cannot occur.

Both versions of the original market can be compared with the same reference market in which $j$ is unavailable. Since each version differs from the reference market in the booking status of at most one customer, the two versions differ in the booking statuses of at most two customers. Similarly, the final set of available listings in each version differs from that of the reference market by at most one listing. Thus, the two versions differ in the booking statuses of at most two listings.

Now suppose that customer $i$ is changed. Compare either version with
the market that skips customer $i$ in the arrival order. The histories or arrivals and bookings coincide before
$i$ arrives. Immediately after that arrival, the market containing
$i$ has either the same available set or one fewer available listing.
The same induction as when we excluded one listing, with the inclusion reversed, shows that among
customers other than $i$, at most one additional customer can fail
to book when $i$ is present. The difference in final listing booking
indicators is at most one for each comparison with the market
without $i$. If $i$ books in both versions, the two possibly
displaced customers are the only customers whose booking statuses
can differ; if $i$ books in just one version, at most $i$ and one
displaced customer can differ. Thus there are at most two different
customer match statuses and at most two different listing match statuses.

Finally, if a changed customer never books in either version of the market, its
arrival does not change the available set, and all subsequent
choices agree. Similarly, suppose listing \(j\) is not booked in either version of the market. If the two booking histories differed between the two markets, consider the first customer whose booking decision differs. At that arrival, the same listings are available in both markets, and customer preferences over all listings other than \(j\) agree. Therefore, any difference in the customer's booking decision must involve booking \(j\) in at least one market, contradicting the assumption. Hence the two booking histories are identical.
\end{proof}

\begin{lemma}[Random experiment group sizes]
\label{lem:fcfs-variance-random-denominators}
Suppose $\varepsilon\le\pi_d\le1-\varepsilon$ and $N_d\to\infty$.
With the convention in 
\eqref{eq:fcfs-difference-in-means} for handling empty expreiment groups, there is a constant
$C_\varepsilon<\infty$ such that
\begin{equation}
    \mathbb E\left[
        \bigl(n\widehat\Delta_{d,n}-H_{d,n}\bigr)^2
    \right]
    \le C_\varepsilon\frac{m_n^2}{N_d}.
    \label{eq:fcfs-variance-dm-HT-bound}
\end{equation}
In particular, for long-side randomization with
$m_n/\ell_n\to0$, the right-hand side is $o(m_n)$.
\end{lemma}

\begin{proof}
Suppress $d,n$. Write $N=N_d$, $m=m_n$, $\pi=\pi_d$, and
$T=N_{d,1}\sim\operatorname{Binomial}(N,\pi)$. On the event
$\mathcal B=\{|T-\pi N|\le\varepsilon N/2\}$,
both experiment groups are nonempty, and
\begin{align*}
    |n\widehat\Delta-H|
    &\le
       K_{d,1}\left|\frac NT-\frac1\pi\right|
       +K_{d,0}\left|\frac N{N-T}-\frac1{1-\pi}\right|\\
    &\le C_\varepsilon(K_{d,1}+K_{d,0})
           \frac{|T-\pi N|}{N}\\
    &\le C_\varepsilon m\frac{|T-\pi N|}{N}.
\end{align*}
The first two inequalities are deterministic and do not require
independence between the size of the experiment groups and the match counts.
Since $\mathbb E[(T-\pi N)^2]=N\pi(1-\pi)$,
\[
    \mathbb E\bigl[
       (n\widehat\Delta-H)^2\mathbf 1_{\mathcal B}
    \bigr]
    \le C_\varepsilon m^2/N.
\]
On $\mathcal B^c$, the rescaled difference in means is bounded by
$N$ in absolute value, including when an experiment group is empty, and
$|H|\le m/\varepsilon$. Hoeffding's inequality gives
$\Pr(\mathcal B^c)\le2e^{-c_\varepsilon N}$.
As $1\le m\le N$, the resulting contribution
$C_\varepsilon N^2e^{-c_\varepsilon N}$ is at most
$C_\varepsilon m^2/N$ after increasing the constant if needed.
This proves the claim. When $N=\ell_n$, we have
$m_n^2/\ell_n=o(m_n)$.
\end{proof}

\begin{proof}[Proof of Proposition~\ref{prop:fcfs-variance-upper}]
For each participant $v$, collect its random primitives and, if it is
on the randomized side, its treatment assignment into one block
$X_v$. These $n$ blocks are independent under the setup of Section~\ref{sec:model}.
Let $X_v'$ be an independent copy of $X_v$, and let $H_{d,n}^{(v)}$ denote
the resulting scaled HT estimator \eqref{eq:fcfs-variance-HT} after replacing $X_v$ with $X_v'$.

By Lemma~\ref{lem:fcfs-variance-stability}, replacing one block changes the booking statuses of at most two participants on the randomized side. If \(v\) is itself on the randomized side, its treatment assignment may also change, affecting only its own coefficient in \(H_{d,n}\) as it gets included in a different experiment group. Since the coefficients in the scaled HT estimator are bounded in absolute value by \(1/\varepsilon\), there exists a constant \(C_\varepsilon\) such that
\[
|H_{d,n}-H_{d,n}^{(v)}|\le C_\varepsilon.
\]
Furthermore, the lemma establishes that if \(v\) is unmatched in both the original and resampled markets, the booking histories are identical. In this case, the scaled HT estimator is unchanged, even if \(v\)'s treatment assignment changes. Combining these two observations and enlarging \(C_\varepsilon\) if necessary gives
\[
(H_{d,n}-H_{d,n}^{(v)})^2
\le C_\varepsilon
\mathbf 1\{v\text{ is matched in at least one of the two markets}\}.
\]
Taking expectations and applying the union bound,
\[
\mathbb E[(H_{d,n}-H_{d,n}^{(v)})^2]
\le C_\varepsilon\bigl\{
\Pr(v\text{ matched})
+\Pr(v\text{ matched after resampling})
\bigr\}.
\]
We now apply the Efron--Stein inequality to obtain
\[
\begin{aligned}
\operatorname{Var}(H_{d,n})
&\le \frac12\sum_v
\mathbb E[(H_{d,n}-H_{d,n}^{(v)})^2]\\
&\le C_\varepsilon\sum_v
\bigl\{\Pr(v\text{ matched})
+\Pr(v\text{ matched after resampling})\bigr\}\\
&=4C_\varepsilon\mathbb E[K_n^d]
\le 4C_\varepsilon m_n.
\end{aligned}
\]
The equality follows because each booking matches one customer and one listing, so the sum of matching probabilities across all participants equals twice the expected number of bookings. The same holds after resampling, since the resampled market has the same distribution as the original market.

Lemma~\ref{lem:fcfs-variance-random-denominators} now implies
\begin{align*}
    \operatorname{Var}(n\widehat\Delta_{d,n})
    &\le2\operatorname{Var}(H_{d,n})
       +2\mathbb E[
           (n\widehat\Delta_{d,n}-H_{d,n})^2]\\
    &\le C_\varepsilon\left(m_n+\frac{m_n^2}{N_d}\right)
     =O(m_n),
\end{align*}
since $N_d\ge m_n$. Dividing by $n^2$ proves the proposition.
\end{proof}

\begin{proof}[Proof of Theorem~\ref{thm:fcfs-variance-long-lower}:
customers scarce]
Suppose $N_C=m_n$ and $N_P=\ell_n$, and write $m=m_n$ and
$\ell=\ell_n$. Condition on $\mathcal E_n$, which contains
the listing primitives and their treatment labels. Set
$a_1=1/\pi_P$ and $a_0=-1/(1-\pi_P)$.
For each customer define the contribution of each customer to the HT estimator in the isolated market as
\[
    G_i=
    \begin{cases}
       a_1,&J_i(\mathcal P_n)\in\mathcal P_{n,1},\\
       a_0,&J_i(\mathcal P_n)\in\mathcal P_{n,0},\\
       0,&J_i(\mathcal P_n)=\varnothing.
    \end{cases}
\]
The $G_i$ are independent conditional on $\mathcal E_n$,
because the customer inputs are independent. For a customer
satisfying \eqref{eq:fcfs-variance-choice-overlap}, write
$p_{i1}=\Pr(G_i=a_1\mid\mathcal E_n)$ and
$p_{i0}=\Pr(G_i=a_0\mid\mathcal E_n)$.
The variance identity for a finite-valued random variable gives
\[
    \operatorname{Var}(G_i\mid\mathcal E_n)
    \ge p_{i1}p_{i0}(a_1-a_0)^2
    \ge\eta^2(a_1-a_0)^2.
\]
Consequently, on the event that at least $\alpha m$ customers
satisfy \eqref{eq:fcfs-variance-choice-overlap},
\begin{equation}
    \operatorname{Var}\left(
        \sum_{i=1}^{m}G_i\,\middle|\,\mathcal E_n
    \right)\ge c m.
    \label{eq:fcfs-variance-isolated-lower}
\end{equation}
The event has probability tending to one.

Write
\[
    H_{L,n}=\sum_{i=1}^{m}G_i+R_n,
\]
where $R_n$ is the correction due to competition for previously
booked listings. We establish
\begin{equation}
    \operatorname{Var}(R_n\mid\mathcal E_n)
    \le C\frac{m^2}{\ell}.
    \label{eq:fcfs-variance-competition-correction}
\end{equation}

To establish this bound, independently resample the full input of
customer $i$, keeping the other customer inputs and $\mathcal E_n$
fixed, and write $R_n^{(i)}$ for the resulting correction. We first bound the probability that customer $i$'s first choice is booked in the original or sampled markets, then on the event that customer $i$ is able to book their first choice in both markets we bound the probability that a subsequent customer's choices differ. We finally show that these are the only ways in which resampling customer $i$'s inputs could change the competition correction and use this to bound the probability that the competition corrections differ.

We first bound the probability that customer $i$ is unable to book their first choice in either market. First, note that the original and resampled booking histories agree before customer
$i$ arrives. Since at most $m$ listings have been booked by that
point, \eqref{eq:fcfs-variance-uniform-dispersion} and independence
of customer $i$'s primitives from preceding customers imply that
their isolated first choice is already booked with probability
at most $Dm/\ell$. The same bound applies to their first choice
under the resampled primitives. By a union bound, the probability
that either first choice is unavailable is at most $2Dm/\ell$.

Now on the complement of these two events, customer $i$ books its first choice in both histories, including choosing the outside option
if it is preferred. Thus, immediately after customer $i$ arrives,
the available sets in the two histories differ by at most the
two listings chosen under the original and resampled primitives. Now consider subsequent customers. As long as a customer makes
the same choice in both histories, the same listing is removed
from both available sets (or neither set changes if the customer
chooses the outside option). Consequently, until the first
subsequent customer whose choices differ, the symmetric difference
between the available sets contains at most two listings. 

Let customer $k>i$ be the first subsequent customer whose choices
differ, if such a customer exists. Since customer $k$ has the
same preferences in both histories, choosing differently requires
selecting a listing belonging to the
symmetric difference of the available sets. There are at most
two such listings. Conditional on the preceding histories and
$\mathcal E_n$, customer $k$'s primitives are independent of
those of preceding customers, so
\eqref{eq:fcfs-variance-uniform-dispersion} bounds the probability
of this event by $2D/\ell$. Applying a union bound over at most
$m$ possible first subsequent differences gives a probability
of at most $2Dm/\ell$ that any subsequent customer's choices
differ.

Now we claim that if the competition corrections differ, either customer $i$ was unable to book their first choice in one of the two markets or some subsequent customer's choices differ. Indeed, if customer $i$ does not encounter their first
choice getting booked in either market and no subsequent customer makes different choices,
then the change in $H_{L,n}$ is exactly $G_i-G_i^{(i)}$:
customer $i$ makes their first choice from the isolated market in both histories
and all other customers' actual and isolated choices remain
unchanged. Hence $R_n=R_n^{(i)}$ on this event. Combining the
two probability bounds thus yields
\[
    \Pr(R_n\ne R_n^{(i)}\mid\mathcal E_n)
    \le C\frac{m}{\ell}.
\]

By Lemma~\ref{lem:fcfs-variance-stability}, the change in
$H_{L,n}$ under resampling is uniformly bounded. Since
$G_i$ and $G_i^{(i)}$ are also uniformly bounded, we have
$|R_n-R_n^{(i)}|\le C_\varepsilon$. Therefore,
\[
    \mathbb E[(R_n-R_n^{(i)})^2\mid\mathcal E_n]
    \le C\frac{m}{\ell}.
\]
Applying the conditional Efron--Stein inequality gives
\[
    \operatorname{Var}(R_n\mid\mathcal E_n)
    \le\frac12\sum_{i=1}^{m}
       \mathbb E[(R_n-R_n^{(i)})^2\mid\mathcal E_n]
    \le C\frac{m^2}{\ell}.
\]
This proves \eqref{eq:fcfs-variance-competition-correction}.

On the event where \eqref{eq:fcfs-variance-choice-overlap} holds, the reverse triangle inequality for
conditional standard deviations now gives
\[
    \sqrt{\operatorname{Var}(H_{L,n}\mid\mathcal E_n)}
    \ge\sqrt{cm}-\sqrt{Cm^2/\ell}.
\]
Since $m/\ell\to0$, this is at least $c'\sqrt m$ for
sufficiently large $n$. The event \eqref{eq:fcfs-variance-choice-overlap} has probability
tending to one, so the law of total variance implies
$\operatorname{Var}(H_{L,n})\ge c''m$.
Finally,
Lemma~\ref{lem:fcfs-variance-random-denominators} gives
$\mathbb E[(n\widehat\Delta_{L,n}-H_{L,n})^2]=o(m)$.
The reverse triangle inequality for unconditional standard
deviations then proves
$\operatorname{Var}(\widehat\Delta_{L,n})
  =\Omega(m/n^2)$.
\end{proof}

\begin{proof}[Proof of Theorem~\ref{thm:fcfs-variance-long-lower}:
listings scarce]
Suppose $N_P=m_n$ and $N_C=\ell_n$. Write $m=m_n$,
$\ell=\ell_n$, and $\pi=\pi_C$. Let $\mathcal E_n$ contain
all listing inputs, and condition on $\mathcal E_n$ throughout
the next argument. For customer $i$, let $W_i\in\{0,1\}$ be
its independent treatment label, and let
$J_i\in\mathcal P_n\cup\{\varnothing\}$ record its actual
booking choice.

Consider the complete booking trajectory
$\mathcal T=(J_1,\ldots,J_\ell)$. For any realized trajectory
$t=(j_1,\ldots,j_\ell)$, the available set immediately
before customer $i$ is a deterministic function $A_i(t)$
of the earlier choices. In particular, it does not depend
on treatment assignments once the trajectory is specified.
Because the customer inputs and treatment assignments are
independent across arrivals,
\begin{equation}
    \Pr(W=w,\mathcal T=t\mid\mathcal E_n)
    =\prod_{i=1}^{\ell}
       \pi^{w_i}(1-\pi)^{1-w_i}
       Q_i^{w_i}(j_i\mid A_i(t)).
    \label{eq:fcfs-variance-trajectory-factorization}
\end{equation}
Here $Q_i^w(\varnothing\mid A)$ denotes the probability of
no booking. The factorization applies to any trajectory of
positive probability.

Bayes' rule applied to
\eqref{eq:fcfs-variance-trajectory-factorization} shows
that conditional on $(\mathcal T,\mathcal E_n)$, the labels
$W_i$ are independent Bernoulli variables with probabilities

\begin{equation}
    \rho_i(\mathcal T)
    =
    \frac{\pi Q_i^1(J_i\mid A_i)}
    {\pi Q_i^1(J_i\mid A_i)
     +(1-\pi)Q_i^0(J_i\mid A_i)}.
    \label{eq:fcfs-variance-booked-label-posterior}
\end{equation}
For every realized booking, the probability of making that choice, averaged over the two treatment assignments, is strictly positive. Under
\eqref{eq:fcfs-variance-booking-overlap}, either both booking
probabilities are positive or neither is, so their ratio lies
between $M^{-1}$ and $M$ on every realized booking.
Since $\varepsilon\le\pi\le1-\varepsilon$, there is a
constant $\eta>0$, independent of $n$, such that for all
customers with $J_i\ne\varnothing$,
\begin{equation}
    \eta\le\rho_i(\mathcal T)\le1-\eta.
    \label{eq:fcfs-variance-booked-label-overlap}
\end{equation}
No such bound is needed for customers who do not book.

Recall that
\[
    H_{L,n}
    =\sum_{i=1}^{\ell}
       \mathbf1\{J_i\ne\varnothing\}
       \left(\frac{W_i}{\pi}
             -\frac{1-W_i}{1-\pi}\right).
\]
The booking indicators are fixed by $\mathcal T$. Since the treatment assignments are independent conditional on \((T,\mathcal E_n)\), we have
\begin{align}
    \operatorname{Var}(H_{L,n}\mid\mathcal T,\mathcal E_n)
    &=\frac1{\pi^2(1-\pi)^2}
      \sum_{i:J_i\ne\varnothing}
          \rho_i(\mathcal T)(1-\rho_i(\mathcal T))
          \notag\\
    &\ge cK_n^L.
    \label{eq:fcfs-variance-trajectory-lower}
\end{align}
The law of total variance and
\eqref{eq:fcfs-variance-positive-utilization} now imply
\[
    \operatorname{Var}(H_{L,n})
    \ge\mathbb E[
       \operatorname{Var}(H_{L,n}\mid\mathcal T,\mathcal E_n)]
    \ge c\mathbb E[K_n^L]
    \ge c\kappa m.
\]
Finally,
Lemma~\ref{lem:fcfs-variance-random-denominators} gives
$\mathbb E[(n\widehat\Delta_{L,n}-H_{L,n})^2]
  \le Cm^2/\ell=o(m)$.
The reverse triangle inequality for standard deviations
therefore gives
$\operatorname{Var}(\widehat\Delta_{L,n})
  =\Omega(m/n^2)$.
\end{proof}

\begin{proof}[Proof of Corollary~\ref{cor:fcfs-variance-logit}]
We first translate the setup of the example into the framework introduced in Section~\ref{sec:model}. Give customer $i$
independent standard Gumbel shocks $G_{ij}$ for the listings
and $G_{i0}$ for the outside option, and generate each
potential consideration indicator $C_{ij}(w)$ from
customer-specific random inputs. When $C_{ij}(w)=1$, set
\[
    u_{ij}(w)=\log v_{ij}^w+G_{ij}
              -\log(\eta_iN_P)-G_{i0};
\]
when $C_{ij}(w)=0$, set $u_{ij}(w)=-1$.
Use the same shocks and consideration inputs across potential
treatment exposure regimes. The logit probabilities then arise from
maximizing these utilities over the considered available
listings and the outside option, and each customer has a
fixed ranking across available sets. The random inputs
are independent across customers and local to the pair
utilities, as required in Section~\ref{sec:model}.

Write
\[
    c_-=
    \frac{\underline p\,\underline v}
         {\overline\eta+\overline v},
    \qquad
    c_+=\frac{\overline v}{\underline\eta}.
\]
Fix a customer $i$, an available set $A$, and a treatment exposure
vector $\mathbf w$. For any $j\in A$, the probability of
booking $j$, after averaging over consideration and logit
choice, is
\[
    \mathbb E\left[
        \frac{C_{ij}(w_{ij})v_{ij}^{w_{ij}}}
        {\eta_iN_P+
         \sum_{k\in A}C_{ik}(w_{ik})v_{ik}^{w_{ik}}}
    \right].
\]
The denominator is always between
$\underline\eta N_P$ and
$(\overline\eta+\overline v)N_P$; the numerator
is positive when $j$ is considered, an event of
probability at least $\underline p$. Therefore
\begin{equation}
    \frac{c_-}{N_P}
    \le
    \Pr(J_i(A)=j\mid\mathbf w,\text{ listing inputs})
    \le\frac{c_+}{N_P}
    \qquad(j\in A).
    \label{eq:fcfs-variance-heterogeneous-choice-bounds}
\end{equation}
The same argument holds conditional on the listing inputs
when the parameters depend on independent participant
characteristics and the stated bounds hold almost surely.
Independence of consideration indicators within a customer
is not needed for this inequality.

First suppose customers are scarce, so $N_P=\ell_n$.
The upper bound in
\eqref{eq:fcfs-variance-heterogeneous-choice-bounds}
verifies \eqref{eq:fcfs-variance-uniform-dispersion}
for every available set. By concentration of the independent
listing assignments, with probability tending to one
both experiment groups contain at least
$\varepsilon\ell_n/2$ listings. On that event,
\eqref{eq:fcfs-variance-heterogeneous-choice-bounds}
implies for every customer $i$ and both $w\in\{0,1\}$,
\[
    \Pr(J_i(\mathcal P_n)\in\mathcal P_{n,w}
         \mid\mathcal E_n)
    \ge\frac{c_-}{\ell_n}|\mathcal P_{n,w}|
    \ge\frac{c_-\varepsilon}{2}.
\]
Thus, both the bounded choice probability and overlap conditions in Assumption~\ref{ass:fcfs-variance-customers-scarce} are satisfied.

Now suppose listings are scarce, so $N_P=m_n=m$ and
$N_C=\ell_n=\ell$. Under customer randomization, for
each available listing $j$ and treatment state $w$,
\[
    \frac{c_-}{m}
    \le Q_i^w(j\mid A)
    \le\frac{c_+}{m}.
\]
Thus
\eqref{eq:fcfs-variance-booking-overlap} holds with
$M=c_+/c_-$. (The constants satisfy $c_+\ge c_-$
because they bound the same nonzero choice probabilities.)
Moreover, if $r$ listings remain just before an arrival,
the conditional booking probability is at least $c_-r/m$.
If $R_t$ is the number of listings remaining after $t$
customer arrivals, independence of the arriving customer's
input from the preceding history gives
\[
    \mathbb E[R_{t+1}\mid R_t=r]
    \le r\left(1-\frac{c_-}{m}\right).
\]
Iterating from $R_0=m$ yields
\[
    \mathbb E[m-K_n^L]
    =\mathbb E[R_{\ell}]
    \le m\left(1-\frac{c_-}{m}\right)^{\ell}
    \le m\exp\left(-c_-\frac{\ell}{m}\right)
    =o(m).
\]
In particular, $\mathbb E[K_n^L]\ge\kappa m$
for any fixed $\kappa<1$ and all sufficiently large $n$.
Thus, both the bounded treatment effect on booking probabilities and sufficient matching conditions in Assumption~\ref{ass:fcfs-variance-listings-scarce} are satisfied. Applying
Theorem~\ref{thm:fcfs-variance-long-lower} and
Proposition~\ref{prop:fcfs-variance-upper} proves the corollary.
\end{proof}

More generally, the bounded likelihood-ratio condition in
Assumption~\ref{ass:fcfs-variance-listings-scarce} does not require uniformly bounded baseline
utility weights. For ordinary logit choice with
\[
    Q_i^w(j\mid A)
    =\frac{v_{ij}^w}
           {v_{i0}^w+\sum_{k\in A}v_{ik}^w},
\]
it suffices that there exist constants $0<a\le b<\infty$
such that $v_{ij}^1/v_{ij}^0\in[a,b]$ for every customer,
listing, and outside option. Indeed, this implies
\[
    \frac ab
    \le\frac{Q_i^1(j\mid A)}{Q_i^0(j\mid A)}
    \le\frac ba.
\]
Hence, the bounded likelihood-ratio condition in Assumption~\ref{ass:fcfs-variance-listings-scarce} holds.

\end{document}